\documentclass[review]{elsarticle}

\usepackage{lineno,hyperref}
\usepackage{tikz-cd}
\usepackage{dsfont}
\usepackage{amsfonts}
\usepackage{amsmath}
\usepackage{slashed}
\usepackage{graphics}
\usepackage{amsmath}
\usepackage{amssymb}
\usepackage{tikz-cd}
\usepackage[compat=1.1.0]{tikz-feynman}
\usepackage{slashed}
\usepackage{setspace}
\usepackage{float}
\modulolinenumbers[5]

\journal{...}
\begin{document}

\begin{frontmatter}

\title{\textbf{Predictions of Ultra-High Energy Cosmic Ray Propagation in the Context of Homogeneously Modified Special Relativity}}

\author[Unimi]{M.D.C. Torri\corref{mycorrespondingauthor}}
\ead{marco.torri@unimi.it - marco.torri@mi.infn.it}

\author[Unimi]{L. Caccianiga}

\author[Unito]{A. di Matteo}

\author[Unimi]{A. Maino}

\author[Unimi]{L. Miramonti}

\address[Unimi]{Dipartimento di Fisica, Universit\'a degli Studi e INFN\\
               via Celoria 16, 20133, Milano, Italy}

\address[Unito]{INFN, Sezione di Torino,\\
               via Pietro Giuria 1, 10125 Torino, Italy}

\begin{abstract}
Ultra high energy cosmic rays (UHECRs) may interact with photon backgrounds and thus the universe is opaque to their propagation. Many Lorentz Invariance Violation (LIV) theories predict a dilation of the expected horizon from which UHECRs can arrive to Earth, in some case even making the interaction probability negligible. In this work, we investigate this effect in the context of the LIV theory that goes by the name of \textit{Homogeneously Modified Special Relativity} (HMSR). In this work, making use of a specifically modified version of the \textit{SimProp} simulation program in order to account for the modifications introduced by the theory to the propagation of particles, the radius of the proton opacity horizon (GZK sphere), and the attenuation length for the photopion production process are simulated and the modifications of these quantities introduced by the theory are studied.
\end{abstract}


\end{frontmatter}

\section{Introduction}
Cosmic rays (CRs) are the part of the radiation of extraterrestrial origin made up by charged particles. In~this work, we deal only with ultra-high energy cosmic rays (UHECRs), which are made up by protons and other atomic nuclei with energies above 1~EeV most likely of extraterrestrial (most likely extragalactic~\cite{Aab:2017tyv,Abreu:2012ybu,Tinyakov:2015qfz,Abbasi:2016kgr,Blasi:2014roa}) origin.  They can be classified based on their composition in three main groups: the~light component principally composed by protons (e.g., protons and He bare nuclei), an~intermediate group (e.g., C, N and O bare nuclei), and the heavy component composed of iron type bare nuclei (\mbox{e.g., Fe and Ni}).

Nowadays, it is recognized that the Universe is opaque to the propagation of cosmic rays (CRs) with energy larger than $\sim$$5\times 10^{19}\,\mathrm{eV}$; this means that such high energy particles can have only a finite free propagation path. Indeed, the propagation of these particles is influenced by their interactions with the \textit{Cosmic Microwave Background} (CMB). Through these processes, UHECRs dissipate energy during their propagation so that they are attenuated in a way that depends on their energy and nature. This~means that UHECRs have a finite free path of order $100\, \mathrm{Mpc}$ for $10^{20}\, \mathrm{eV}$ protons and even shorter for higher energies or heavier nuclei, hence, if they propagate for longer distances, they can be detected on Earth only under a determined energy threshold. This phenomenon is named GZK-cutoff after the physicists Greisen, Zatsepin, and Kuzmin, who first theorized it~\cite{Greisen,Zatsepin}, posing an upper limit on the energy of UHECR (which are supposed of extragalactic origin) that can be detected on~Earth.

The UHECR sources are still unknown; nevertheless, the physics correlated with their propagation can be exploited to investigate new phenomena, such as the hypothetical ones induced by quantum gravity and described by LIV theories. These perturbations can be considered as Planck scale physics residual effects in the low energy limit; therefore, they may be an open window on possible phenomenology induced by quantum gravity. These effects can be considered caused by the interaction of particles with the presumed quantum structure of the background~spacetime.

In this work, we introduce a new methodology to investigate this sector exploiting the peculiar features of UHECR physics. Indeed, the CR opacity horizon may be enlarged by new physics, such as the introduction of LIV extensions to the particle Standard Model (SM) \cite{Coleman, Scully, Stecker, TorriUHECR, TorriPhD}.

In this study, to~investigate LIV effects on UHECR propagation, we limit our analysis to the lightest component of UHECRs, which is protons, since the lightest component of UHECRs is the least deflected by magnetic fields and propagates on more straightforward trajectories than heavier nuclei. The~LIV effects are expected to be more visible for the intermediate component of CRs and composition data show a mix of light and intermediate elements (except possibly at the highest energies) so all the constraints on LIV obtained for the light component are still valid for the real physical case, for~which they are expected to be even more~stringent.

There are various LIV theoretical models~\cite{Koste1,Coleman,Cohen,AmelinoCamelia,AmelinoCamelia2,AmelinoCamelia3,AmelinoCamelia4}, but~we resort to \emph{Homogeneously Modified Special Relativity} (HMSR) \cite{TorriHMSR} since, for this kind of analysis, it is useful that the theory preserves spacetime isotropy and essential that it can foresee an extension of the particle Standard Model. The~introduction of LIV in UHECR physics can determine a kinematical perturbation that modifies the allowed phase space of the photopion production process. The~perturbation is then introduced in the \textit{SimProp} \cite{Aloisio} software that becomes able to simulate the LIV modified propagation of light cosmic rays in order to evaluate the GZK sphere radius modifications and the UHECR attenuation~length.

This work is structured as follows: in Section~\ref{sec:HMSR}, we present the LIV model employed here (HMSR)~\cite{TorriHMSR}; in Section~\ref{sec:GZK}, we introduce UHECR propagation; in Section~\ref{sec:propagation}, we illustrate the modifications introduced in the UHECR physics by LIV in order to obtain a modified  \textit{SimProp} software version able to simulate their free propagation in a Lorentz violating scenario; finally, in~Section~\ref{sec:simulations}, we summarize the obtained results and, in Section~\ref{sec:conclusions}, we present the future perspectives to extend the analysis to a more realistic CR composition~scenario.


\section{Homogeneously Modified Special~Relativity}\label{sec:HMSR}
In this work, we explore the possibility to introduce LIV to justify a possible GZK phenomenon modification, in~order to explore quantum gravity effects. The~purpose of our study implies the necessity to use a LIV model that can both modify particle physics phenomenology as required and preserve the isotropy of spacetime. We decided to set our study in the HMSR theoretical framework~\cite{TorriHMSR}, since this model was developed in order to investigate the opportunity to extend the Standard Model (SM) of elementary particles preserving spacetime isotropy and homogeneity in a LIV~scenario.

\subsection{Geometric Structure: Hamilton/Finsler~Geometry}
In order to obtain an extension of the particle SM that preserves spacetime isotropy and does not introduce any exotic reaction, only the free propagation of massive fermions is modified geometrizing their interaction with the assumed background quantum structure. This approach is exploited modifying the dispersion relations (DR) of massive particles, as~in the majority of LIV theories~\cite{Koste1,Coleman,Cohen,AmelinoCamelia,AmelinoCamelia2,AmelinoCamelia3,AmelinoCamelia4,Smolin1,Smolin2}.

Every particle species is supposed to have a personal modified dispersion relation (MDR) with the form:
\begin{equation}
\label{a1}
\text{MDR}\,\Rightarrow\,F^{2}(\mathbf{p})=E^2-\left|\vec{p}\right|^2\left(1-f\left(\frac{\left|\vec{p}\right|}{E}\right)\right)=m^2
\end{equation}
where the bold letters indicate four-vectors; the~perturbation function $f$ is related to the particle species and depends only on the magnitude of the three-momentum vector $|\vec{p}|$ in order to be rotationally invariant. The~perturbation function is chosen homogeneous of degree $0$ to preserve the geometrical origin of the MDR. In~this way, the $F^2(\mathbf{p})$ function is homogeneous of degree $2$ and can be a candidate Finsler pseudo-norm. Here, we are dealing with pseudo-Finsler structures, since the local underlying geometry is the Minkowski one, with~metric $\eta_{\mu\nu}=\operatorname{diag}(1,\,-1,\,-1,\,-1)$ that is not positive definite. On~the contrary, a Finsler geometry is constructed starting from a positive defined metric. Some issues can emerge in defining a pseudo-Finsler structure~\cite{Pfeifer1,Pfeifer2,Pfeifer3,Javaloyes,Bernal}, but,~in the case of HMSR, the function $f$ is supposed to be of a tiny magnitude compared to the other MDR quantities, so it represents a perturbation; therefore, as a result, it is possible to deal with these issues. In~the actual HMSR formulation, MDR are supposed CPT even, since they do not present an explicit dependence on particle helicity or spin. Indeed as a result, it is possible to introduce LIV preserving the CPT symmetry. It is well known that Lorentz violation does not imply the CPT one~\cite{Greenberg1,Greenberg2,TorriCPT}, whereas the opposite statement is widely debated in literature~\cite{Discussion1,Discussion2,Discussion3,Discussion4,Discussion5}.

In HMSR, every particle species has its own personal MDR, with~a specific perturbation $f$. This~function depends on the ratio $|\vec{p}|/E$ which admits a finite limit:
\begin{equation}
\label{a2}
\begin{split}
&\frac{|\vec{p}|}{E}\rightarrow(1+\delta)\;\;\text{for}\;\;E\rightarrow\infty\;\;\text{with}\;\;0<\delta\ll1\\
&f\left(\frac{|\vec{p}|}{E}\right)\rightarrow\epsilon\ll1
\end{split}
\end{equation}

This last relation implies that every massive particle admits a different personal maximum attainable velocity (MAV):
\begin{equation}
\label{a3}
c'=\left(1-f\left(\frac{|\vec{p}|}{E}\right)\right)\rightarrow(1-\epsilon)
\end{equation}
as assumed in the first formulation of Very Special Relativity~\cite{Coleman}. This HMSR feature is of a very general nature and allows for obtaining a rich phenomenology even in the neutrino oscillation physics~\cite{Torrineutrini,Torri2020}.

The MDR (\ref{a1}) can be promoted to the role of momentum space norm, making it possible to obtain the associated momentum space Finsler one, using the equation:
\begin{equation}
\label{a4}
\widetilde{g}(\mathbf{p})^{\mu\nu}=\frac{1}{2}\frac{\partial}{\partial p^{\mu}}\frac{\partial}{\partial p^{\nu}}F^2(E,\,\vec{p})
\end{equation}

The obtained metric presents a non diagonal part that does not contribute to the computation of the MDR, so it can be neglected, thanks to the polarization identity, and~the final form of the obtained metric can be reduced to a pure diagonal form, without~a changing of the basis:
\begin{equation}
\label{a5}
\widetilde{g}^{\mu\nu}(\mathbf{p})=\left(
                             \begin{array}{cc}
                                1 & \vec{0} \\
                                \vec{0}^{t} & -\left(1-f\left(\left|\vec{p}\right|/E\right)\right)\mathbb{I}_{3\times3} \\
                             \end{array}
                          \right)
\end{equation}

Considering the peculiar form of the perturbation function $f$ that depends on the ratio $|\vec{p}|/E$~(\ref{a2}), it is possible to correlate at least at the leading order the momentum space with the coordinate space via the Legendre transform. Indeed, the homogeneity of the perturbation $f$ allows for neglecting the derivatives with respect to the momentum, since these terms are a second order perturbation~\cite{TorriHMSR}. HMSR introduces therefore a simple way of inverting the correspondence between four-momentum and four-velocity.\footnote{the demonstration that derivatives of this homogeneous function are negligible is given in \ref{app}}
 The final form of the coordinate space metric is therefore given by:
\begin{equation}
\label{a6}
g(\mathbf{x},\,\dot{\mathbf{x}}(\mathbf{p}))_{\mu\nu}=\left(
                     \begin{array}{cc}
                         1 & \vec{0} \\
                         \vec{0}^{t} & -\left(1+f\left(\left|\vec{p}\right|/E\right)\right)\mathbb{I}_{3\times3} \\
                     \end{array}
                  \right)\\
\end{equation}

This allows us to correlate the Finsler co-metric resulting from the modified dispersion relation and the Finsler metric describing the modified geometry in coordinate space. The~pseudo-Finsler norm can indeed be written as a function of the coordinates:
\begin{equation}
\label{a7}
G(\dot{\mathbf{x}}(\mathbf{p}))=F(\mathbf{p})
\end{equation}
and even the related metric in the coordinate space can be computed via the relation:
\begin{equation}
\label{a8}
g(\mathbf{x},\,\dot{\mathbf{x}}(\mathbf{p}))_{\mu\nu}=\frac{1}{2}\left(\frac{\partial^2G}{\partial \dot{x}^{\mu}\dot{x}^{\nu}}\right)
\end{equation}

The metric in coordinate space constitutes the inverse of the one in momentum space:
\begin{equation}
\label{a9}
\widetilde{g}^{\mu\alpha}\,g_{\alpha\nu}=\delta^{\mu}_{\,\nu}
\end{equation}

To complete the HMSR modified geometry description, we give the Hamiltonian of the free massive particle as a function of the proper time:
\begin{equation}
\label{a10}
\mathcal{H}=\sqrt{\widetilde{g}^{\mu\nu}(\mathbf{p})\,p_{\mu}\,p_{\nu}}=F(\mathbf{p})
\end{equation}
consistently with standard Special Relativity. The~velocity correlated with the momentum can be computed resorting to the Legendre transformation:
\begin{equation}
\label{a11}
\dot{x}^{\mu}=\frac{\partial}{\partial p_{\mu}}F(\mathbf{p})\simeq\frac{\widetilde{g}^{\mu\nu}(\mathbf{p})\,p_{\nu}}{\sqrt{\widetilde{g}^{\mu\nu}(\mathbf{p})\,p_{\mu}\,p_{\nu}}}=\frac{\widetilde{g}^{\mu\nu}(\mathbf{p})\,p_{\nu}}{m}
\end{equation}

The \emph{Lagrangian} can be evaluated starting from the \emph{Hamiltonian}:
\begin{equation}
\label{a12}
\mathcal{L}=\vec{p}\cdot\dot{\vec{x}}-\mathcal{H}=-\dot{x}^{\mu}\,p_{\mu}=-m\sqrt{g_{\mu\nu}(\mathbf{p})\,\dot{x}^{\mu}\,\dot{x}^{\nu}}
\end{equation}

Starting from the obtained metrics (Equations~(\ref{a5}) and (\ref{a6})), it is possible to introduce the generalized vierbein, in~order to satisfy the definitional equations:
\begin{equation}
\begin{split}
\label{a13}
&g_{\mu\nu}(\dot{\mathbf{x}})=e_{\mu}^{\,a}(\mathbf{p}(\dot{\mathbf{x}}))\,\eta_{ab}\,e_{\nu}^{\,b}(\mathbf{p}(\dot{\mathbf{x}}))\\
&\widetilde{g}^{\mu\nu}(\mathbf{p})=e^{\mu}_{\,a}(\mathbf{p})\,\eta^{ab}\,e^{\nu}_{\,b}(\mathbf{p})
\end{split}
\end{equation}
obtaining the explicit forms:
\begin{equation}
\label{a14}
\begin{split}
& e^{\mu}_{\,a}(\mathbf{p})=        \left(
                             \begin{array}{cc}
                               1 & \vec{0} \\
                               \vec{0}^{t} & \sqrt{1-f\left(\left(\left|\vec{p}\right|/E\right)\right)}\,\mathbb{I}_{3\times3} \\
                             \end{array}
                           \right)\\ \\
& e_{\mu}^{\,a}(\mathbf{p})=        \left(
                             \begin{array}{cc}
                               1 & \vec{0} \\
                               \vec{0}^{t} & \sqrt{1+f\left(\left(\left|\vec{p}\right|/E\right)\right)}\,\mathbb{I}_{3\times3} \\
                             \end{array}
                           \right)
\end{split}
\end{equation}

The peculiar feature of HMSR is that every particle feels a specific personal modified spacetime, parameterized by the particle momentum. Every particle lives in a personal curved spacetime and therefore every physical quantity is generalized, acquiring an explicit dependence on the momentum. It is therefore necessary to introduce an original formalism to correlate different local curved spaces, using the generalized vierbein as a projector from the curved spacetime to a common flat Minkowski support space. Here, we report a scheme of how the original mathematical formalism of HMSR works to relate different local curved spaces:
\vspace{-6pt}

\[\begin{tikzcd}
         (TM,\,\eta_{ab},\,\mathbf{p}) \arrow{d}{e(\mathbf{p})} \arrow{rr}[swap]{\Lambda} && (TM,\,\eta_{ab},\,\mathbf{p}')\arrow{d}[swap]{\overline{e}(\mathbf{p}')} \\
        (T_{x}M,\,g_{\mu\nu}(\mathbf{p})) \arrow{rr}[swap]{\overline{e}\circ\Lambda\circ e^{-1}} && (T_{x}M,\,\overline{g}_{\mu\nu}(\mathbf{p}'))
\end{tikzcd}\]
where the Greek indices refer to the local curved geometric structures, whereas the Latin ones refer to the common support Minkowski space. Using again the vierbein as projectors from the common support space to the local curved one, it is simple to obtain the generalized Lorentz transformations:
\begin{equation}
\label{a15}
\Lambda_{\mu}^{\;\nu}(\mathbf{p})=e_{\;\mu}^{a}(\Lambda\mathbf{p})\,\Lambda_{a}^{\;b}\,e_{\;\nu}^{b}(\mathbf{p})
\end{equation}
where $\Lambda_{a}^{\;b}$ are the classic Lorentz ones valid for the flat Minkowski spacetime. These \emph{Modified Lorentz Transformations} (MLTs) are the isometries of the MDR (\ref{a1}), so every particle species presents its personal MLTs, which preserve its specific~MDR.

The geometrical structure is defined by the \emph{affine} and the \emph{spinorial} connections and here we give their characterization. It is simple to evaluate the explicit forms of the affine connection components, starting from the metric tensor.
The \emph{affine} connection is defined via the \emph{Christoffel} symbols:
\begin{equation}
\label{c28}
\Gamma_{\mu\nu}^{\,\alpha}=\frac{1}{2}g^{\alpha\beta}\left(\partial_{\mu}g_{\beta\nu}+\partial_{\nu}g_{\mu\beta}-\partial_{\beta}g_{\mu\nu}\right)
\end{equation}

The following Christoffel symbols are identically equal to zero:
\begin{equation}
\label{c29}
\Gamma_{\mu0}^{\,0}=\Gamma_{00}^{\,i}=\Gamma_{\mu\nu}^{\,i}=0\qquad\forall \mu\neq\nu
\end{equation}
but even the remaining elements are null at the leading order, since they depend on derivatives of the 0 degree homogeneous perturbation function $f$:
\begin{equation}
\label{c29a}
\begin{split}
&\Gamma_{ii}^{\,0}=-\frac{1}{2}\partial_{0}f(\mathbf{p})\simeq0\\
&\Gamma_{0i}^{\,0}=\Gamma_{i0}^{\,0}=\frac{1}{2(1+f(\mathbf{p}))}\partial_{0}f(\mathbf{p})\simeq0\\
&\Gamma_{ii}^{\,i}=\frac{1}{2(1+f(\mathbf{p}))}\partial_{i}f(\mathbf{p})\simeq0\\
&\Gamma_{jj}^{\,i}=-\frac{1}{2(1+f(\mathbf{p}))}\partial_{i}f(\mathbf{p})\simeq0\qquad\forall i\neq j\\
&\Gamma_{ij}^{\,i}=\Gamma_{ji}^{\,i}=\frac{1}{2(1+f(\mathbf{p}))}\partial_{i}f(\mathbf{p})\simeq0\qquad\forall i\neq j
\end{split}
\end{equation}

The \emph{Cartan} or \emph{spinorial} connection is defined as:
\begin{equation}
\label{c31}
\omega_{\mu ab}=e^{\;\nu}_{a}\nabla_{\mu}e_{b\nu}\simeq e^{\;\nu}_{a}\partial_{\mu}e_{b\nu}
\end{equation}

Even the \emph{Cartan} or \emph{spinorial} connection is negligible at the leading order, since every element is again proportional to derivatives of the 0 degree homogeneous perturbation function $f$. Indeed, from the first Cartan structural equations:
\begin{equation}
\label{c32}
de=e\wedge\omega
\end{equation}
applied to the external forms
\vspace{6pt}
\begin{equation}
\label{c33}
e_{0}^{\,\mu}=dx^{\mu}\qquad e_{i}^{\,\mu}=\sqrt{1-f(\mathbf{p})}dx^{\mu},
\end{equation}
it follows that the spinorial connection elements are given by:
\begin{equation}
\label{c34}
\frac{1}{2}\epsilon_{ijk}\omega^{ij}=\frac{1}{2}\frac{1}{1-f(\mathbf{p})}\epsilon_{ijk}(\partial^{i}f(\mathbf{p})dx^{j}-\partial^{j}f(\mathbf{p})dx^{i})\simeq0
\end{equation}
Since the two fundamental connections can be neglected, the~total geometric covariant derivative becomes:
\begin{equation}
\label{c36}
D_{\mu}v^{\;\nu}_{a}=\partial_{\mu}v^{\;\nu}_{a}+\Gamma_{\mu\alpha}^{\,\nu}v^{\;\alpha}_{a}-\omega_{\mu\nu}^{\,a}v^{\;\nu}_{b}\simeq\partial_{\mu}v^{\;\nu}_{b}
\end{equation}

The resulting geometric structure is an asymptotically flat Finsler pseudo-structure~\cite{Koste11,Liberati1,Edwards,Lammerzahl,Bubuianu,Schreck}.

Even the Poincar\'e brackets of the local coordinates are modified:
{\small
\begin{equation}
\label{a16}
\begin{split}
&\{\widetilde{x}^{\mu},\,\widetilde{x}^{\nu}\}=\{x^{i}e_{i}^{\,\mu}(\mathbf{p}),\,x^{j}e_{j}^{\,\nu}(\mathbf{p})\}=\{x^{i},\,e_{j}^{\,\nu}(\mathbf{p})\}e_{i}^{\,\mu}(\mathbf{p})x^{j}+\{e_{i}^{\,\mu}(\mathbf{p}),\,x^{j}\}x^{i}e_{j}^{\,\nu}(\mathbf{p})=0\\
&\{\widetilde{x}^{\mu},\,\widetilde{p}_{\nu}\}=\{x^{i}e_{i}^{\,\mu}(\mathbf{p}),\,p_{j}e^{j}_{\,\nu}(\mathbf{p})\}=\{x^{i},\,p_{j}\}e_{i}^{\,\mu}(\mathbf{p})e^{j}_{\,\nu}(\mathbf{p})+\{x^{i},\,e^{j}_{\,\nu}(\mathbf{p})\}e_{i}^{\,\mu}(p)p_{j}=\\
&=\delta^{\mu}_{\,\nu}+\{x^{i},\,e^{j}_{\,\nu}(\mathbf{p})\}e_{i}^{\,\mu}(\mathbf{p})p_{j}\\
&\{\widetilde{p}_{\mu},\,\widetilde{p}_{\nu}\}=\{p_{i}e^{i}_{\,\mu}(\mathbf{p}),\,p_{j}e^{j}_{\,\nu}(\mathbf{p})\}=0
\end{split}
\end{equation}}

\noindent where, in the second equation, the parentheses $\{x^{i},\,e^{j}_{\,\nu}(\mathbf{p})\}\neq0$ since the vierbein $e^{j}_{\,\nu}(\mathbf{p})$ is a function of the derivative $p_{i}=\dfrac{\partial}{\partial x^{i}}$. Time and space therefore do not commute anymore~\cite{Torri2020}.

\subsection{HMSR Standard Model~Extension}

In the HMSR scenario, as a result, it is possible to amend the formulation of the SM of particle physics following a procedure analogous to that used in the \emph{Standard Model Extension} (SME) \cite{Koste1}. In~order to modify the theory formulation, it is necessary to start from the definition of the amended Dirac matrices, introducing the usual momentum dependence:
\begin{equation}
\label{a17}
\Gamma^{\mu}=e^{\;\mu}_{a}(\mathbf{p})\,\gamma^{a}\qquad \Gamma_5=\frac{\epsilon^{\mu\nu\alpha\beta}}{4!}\Gamma_{\mu}\Gamma_{\nu}\Gamma_{\alpha}\Gamma_{\beta}=\gamma_5
\end{equation}

Next, it is possible to modify the Clifford Algebra, starting from the anticommutator of the $\Gamma$~matrices:
\begin{equation}
\label{a18}
\{\Gamma_{\mu},\Gamma_{\nu}\}=2\,g^{\mu\nu}(\mathbf{p})=2\,e_{\mu}^{\,a}(\mathbf{p})\,\eta_{ab}\,e_{\nu}^{\,b}(\mathbf{p})
\end{equation}
Finally it is possible to introduce the modified spinor fields, preserving their plane wave formulation:
\begin{equation}
\label{a19}
\begin{split}
&\psi^{+}(x)=u_{r}(\mathbf{p})e^{-ip_{\mu}x^{\mu}}\\
&\psi^{-}(x)=v_{r}(\mathbf{p})e^{ip_{\mu}x^{\mu}}
\end{split}
\end{equation}
where the spinors $u_{r}(\mathbf{p})$ and $v_{r}(\mathbf{p})$ normalization results modified. It can be useful to notice that the modified Dirac equation:
\begin{equation}
\label{a20}
(i\Gamma^{\mu}\partial_{\mu}-m)\psi=0
\end{equation}
implies the MDR (\ref{a1}):
\begin{equation}
\label{a21}
\begin{split}
&(i\Gamma^{\mu}\partial_{\mu}+m)(i\Gamma^{\mu}\partial_{\mu}-m)\psi^{+}=0\,\Rightarrow\,(\Gamma^{\mu}p_{\mu}+m)(\Gamma^{\mu}p_{\mu}-m)u_{r}(\mathbf{p})=0\,\Rightarrow\,\\
\Rightarrow\,&\left(\frac{1}{2}\{\Gamma^{\mu},\,\Gamma^{\nu}\}p_{\mu}p_{\nu}-m^2\right)u_{r}(\mathbf{p})=0\,\Rightarrow\,(p_{\mu}g^{\mu\nu}p_{\nu}-m^2)u_{r}(\mathbf{p})=0
\end{split}
\end{equation}

As a final result, it is possible to formulate the amended particle SM, resorting to the vierbein formalism to project every physical quantity from the modified curved spacetime to the common support flat Minkowski spacetime. For~instance, the quantum electrodynamics (QED) Lagrangian can be written as:
\vspace{6pt}
\begin{equation}
\label{a22}
\mathcal{L}=\sqrt{\left|\det{[g]}\right|}\;\;\overline{\psi}(i\Gamma^{\mu}\partial_{\mu}-m)\psi+e\sqrt{\left|\det{[\widetilde{g}]}\right|}\;\;\overline{\psi}\,\Gamma_{\mu}(\mathbf{p},\,\mathbf{p}')\,\psi\,\overline{e}^{\mu}_{\;\nu}\,A^{\nu}
\end{equation}
where the $\overline{e}$ vierbein has been introduced related to the gauge field, and the index $\mu$ represents a coordinate of the Minkowski space-time $(TM,\,\eta_{\mu\nu})$. In~order to simplify the theory formulation, the gauge bosons can be supposed to be Lorentz invariant. $\sqrt{|\det{[g]}|}$ is the term borrowed from curved space-time~QFT.

While in the low energy scenario the LIV perturbation can be neglected, in~the high energy limit, it is possible to consider incoming and outgoing momenta with approximately the same magnitude, even~after interaction. The~conserved currents therefore admit a high energy constant limit not dependent on the momentum. The~conserved current reduces then to the form:
\begin{equation}
\label{a23}
J_{\mu}=e\sqrt{\left|\det{\tfrac{1}{2}\{\Gamma_{\mu},\,\Gamma_{\nu}\}}\right|}\;\;\overline{\psi}\,\Gamma_{\mu}\,\psi=e\sqrt{\left|\det{[g]}\right|}\;\;\overline{\psi}\,\Gamma_{\mu}\,\psi
\end{equation}

The modified SM formulation preserves the classical gauge $SU(3)\times SU(2)\times U(1)$. The~Coleman--Mandula theorem is still valid in a modified formulation that provides an amended symmetry group given by $\mathcal{P}(\{\mathbf{p}\})\otimes G_{int}$, where $\mathcal{P}(\{\mathbf{p}\})$ is the direct product of modified---particle and momentum depending---Poincar\'e groups:
\begin{equation}
\label{a24}
\mathcal{P}\left(\{\mathbf{p}\}\right)=\bigotimes_{i}\mathcal{P}^{(i)}\bigr(\mathbf{p}_{(i)}\bigr)
\end{equation}
and $G_{int}$ is the internal symmetry group (in this case, $SU(3)\times SU(2)\times U(1)$).

\subsection{HMSR Modified Kinematics}

In HMSR framework, the most interesting phenomenological effects caused by the introduction of LIV  emerge in the interaction of different particle species, since every particle type physics is modified in a different way by LIV.  It is therefore necessary to determine how the reaction invariants---that is, the Mandelstam $s$, $t$, and $u$ variables are modified:
\tikzset{
particle/.style={thick,draw=black, postaction={decorate},
    decoration={markings,mark=at position .5 with {\arrow[black]{triangle 45}}}},
gluon/.style={decorate, draw=black,
    decoration={coil,aspect=0}}
 }
\tikzset{
particle/.style={thin,draw=black, postaction={decorate},
decoration={markings,mark=at position .5 with {\arrow[black]{stealth}}}},
gluon/.style={decorate, draw=black, decoration={snake=coil}}
}
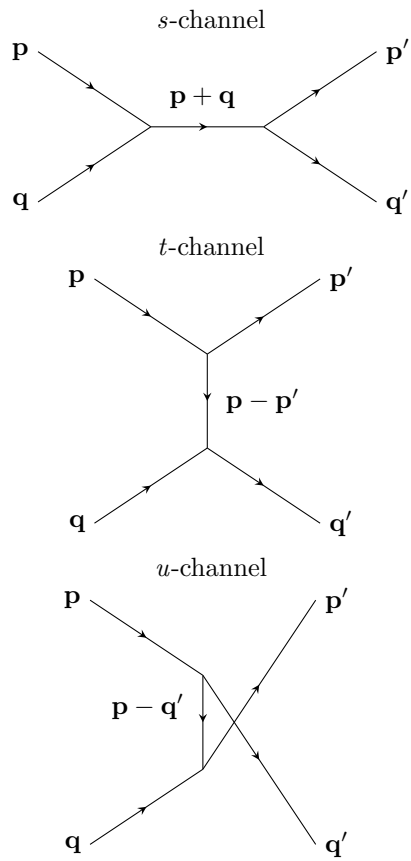
\begin{figure}
\centering
\emph{s}-channel\\

\begin{tikzpicture}[node distance=1cm and 1.5cm]
\coordinate[label=left:$\mathbf{p}$] (e1);
\coordinate[below right=of e1] (aux1);
\coordinate[right=of aux1] (aux2);
\coordinate[above right=of aux2,label=right:$\mathbf{p}'$] (e2);
\coordinate[below left=of aux1,label=left:$\mathbf{q}$] (e3);
\coordinate[below right=of aux2,label=right:$\mathbf{q}'$] (e4);

\draw[particle] (e1) -- (aux1);
\draw[particle] (aux2) -- (e2);
\draw[particle] (e3) -- (aux1);
\draw[particle] (aux2) -- (e4);
\draw[particle] (aux1)  node[label=above right:$\mathbf{p}+\mathbf{q}$] {} -- (aux2);
\end{tikzpicture}

\emph{t}-channel\\

\begin{tikzpicture}[node distance=1cm and 1.5cm]

\coordinate[label=left:$\mathbf{p}$] (e1);
\coordinate[below right=of e1] (aux1);
\coordinate[above right=of aux1,label=right:$\mathbf{p}'$] (e2);
\coordinate[below=1.25cm of aux1] (aux2);
\coordinate[below left=of aux2,label=left:$\mathbf{q}$] (e3);
\coordinate[below right=of aux2,label=right:$\mathbf{q}'$] (e4);

\draw[particle] (e1) -- (aux1);
\draw[particle] (aux1) -- (e2);
\draw[particle] (e3) -- (aux2);
\draw[particle] (aux2) -- (e4);
\draw[particle] (aux1) -- node[label=right:$\mathbf{p}-\mathbf{p}'$] {}(aux2);
\end{tikzpicture}

\newpage

\emph{u}-channel\\
\begin{tikzpicture}[node distance=1cm and 1.5cm]
\coordinate[label=left:$\mathbf{p}$] (e1);
\coordinate[below right=of e1] (aux1);
\coordinate[above right=of aux1,label=right:$\mathbf{p}'$] (e2);
\coordinate[below=1.25cm of aux1] (aux2);
\coordinate[below left=of aux2,label=left:$\mathbf{q}$] (e3);
\coordinate[below right=of aux2,label=right:$\mathbf{q}'$] (e4);

\draw[particle] (e1) -- (aux1);
\draw[particle] (aux2) -- (e2);
\draw[particle] (e3) -- (aux2);
\draw[particle] (aux1) -- (e4);
\draw[particle] (aux1) node[label=below left:$\mathbf{p}-\mathbf{q}'$] {} -- (aux2);
\end{tikzpicture}
\caption{\emph{Feynman diagrams related to Mandelstam variables}}
\end{figure}
considering the $\mathbf{p}$ and $\mathbf{q}$ momenta as belonging in general to different particle species. It is therefore necessary to generalize the internal product definition for the sum of two or more particle species momenta as:
\begin{equation}
\label{a25}
\langle \mathbf{p}+\mathbf{q}|\mathbf{p}+\mathbf{q} \rangle=(p_{\mu}\,e_{a}^{\,\mu}(\mathbf{p})+q_{\mu}\,\tilde{e}_{a}^{\,\mu}(\mathbf{q}))\,\eta^{ab}\,(p_{\nu}\,e_{b}^{\,\nu}(\mathbf{p})+q_{\nu}\,\tilde{e}_{b}^{\,\nu}(\mathbf{q}))
\end{equation}
where the vierbeins $e_{a}^{\,\mu}(\mathbf{p})$ and $\tilde{e}_{a}^{\,\mu}(\mathbf{q}))$---associated respectively with the two different species---are used. Now, one can generalize the definition of the Mandelstam variables $s$, $t$, and $u$ resorting to the modified internal product definition.
The new internal product can be written introducing the generalized metric:
\begin{equation}
\label{c46}
G=\left(
    \begin{array}{cc}
      g^{\mu\nu}(\mathbf{p}) & e^{a\mu}(\mathbf{p})\tilde{e}_{a}^{\;\beta}(\mathbf{q}) \\
      \tilde{e}^{a\alpha}(\mathbf{q})e_{a}^{\;\nu}(\mathbf{p}) & \tilde{g}^{\alpha\beta}(\mathbf{q}) \\
    \end{array}
  \right)
\end{equation}

The internal product (\ref{a25}) can be written as:
\begin{equation}
\begin{split}
\label{c48}
&\langle \mathbf{p}+\mathbf{q}|\mathbf{p}+\mathbf{q} \rangle=
  \begin{pmatrix}
    \mathbf{p} & \mathbf{q} \\
  \end{pmatrix}
\left(
    \begin{array}{cc}
      g^{\mu\nu}(\mathbf{p}) & e^{a\mu}(\mathbf{p})\tilde{e}_{a}^{\;\beta}(\mathbf{q}) \\
      \tilde{e}^{a\alpha}(\mathbf{q})e_{a}^{\;\nu}(\mathbf{p}) & \tilde{g}^{\alpha\beta}(\mathbf{q}) \\
    \end{array}
\right)
  \begin{pmatrix}
    \mathbf{p} \\
    \mathbf{q} \\
  \end{pmatrix}
=\\
=&p_{\mu}\,g^{\mu\nu}(\mathbf{p})\,p_{\nu}+p_{\mu}\, e^{a\mu}\tilde{e}_{a}^{\;\beta}(\mathbf{q})\,q_{\beta}+q_{\alpha}\,\tilde{e}^{a\alpha}(\mathbf{q})\,e_{a}^{\;\nu}(\mathbf{p})\,p_{\nu}+q_{\alpha}\tilde{g}^{\alpha\beta}(\mathbf{q})\,q_{\beta}
\end{split}
\end{equation}

This generalization presents the feature to be invariant respect to the \emph{Modified Lorentz Transfomations} MLTs, that is, HMSR presents a covariant formulation, with respect to the MLTs, with~the explicit form:
\begin{equation}
\label{a28}
\Lambda=
\left(
  \begin{array}{cc}
    \Lambda_{\mu}^{\;\mu'} & 0 \\
    0 & \tilde{\Lambda}_{\alpha}^{\;\alpha'} \\
  \end{array}
\right)
\end{equation}
using the MLTs of the two particle species:
\begin{equation}
\begin{split}
\label{a29}
&\langle \mathbf{p}+\mathbf{q}|\mathbf{p}+\mathbf{q} \rangle=
  \begin{pmatrix}
    \mathbf{p} \\
    \mathbf{q} \\
  \end{pmatrix}
^{t}\cdot G\cdot
  \begin{pmatrix}
    \mathbf{p} \\
    \mathbf{q} \\
  \end{pmatrix}
=\\
=&\left(\Lambda
  \begin{pmatrix}
    \mathbf{p} \\
    \mathbf{q} \\
  \end{pmatrix}
\right)^{t}\cdot \Lambda \cdot G\cdot \Lambda^{t}\cdot\Lambda
  \begin{pmatrix}
    \mathbf{p} \\
    \mathbf{q} \\
  \end{pmatrix}
=\langle \Lambda(\mathbf{p}+\mathbf{q})|\Lambda(\mathbf{p}+\mathbf{q}) \rangle=
\end{split}
\end{equation}
where $\Lambda \cdot G\cdot \Lambda^{t}$ represents the metric evaluated for the two particles momenta given, respectively, by $\Lambda \mathbf{p}$, $\tilde\Lambda \mathbf{q}$. It is therefore clear that HMSR does not require the introduction of a preferred reference frame, in~contrast with most of the other LIV~models.


\section{Ultra High Energy Cosmic Rays Propagation and GZK Cut-Off~Effect}\label{sec:GZK}
UHECRs interact with the backgrounds photons, and they are attenuated in a way that depends on their composition and energy; for~instance, nuclei can undergo a photo-dissociation process
\begin{equation}
\label{1}
A+\gamma\,\rightarrow\,(A-1)+n
\end{equation}
where $A$ is the atomic number of the considered CR bare nucleus, while protons interacting with the background photons can undergo a pair production:
\begin{equation}
\label{2}
p+\gamma\,\rightarrow\,p+e^{-}+e^{+}
\end{equation}
or a photopion process, through a $\Delta$ particle resonance:
\begin{equation}
\label{3}
\begin{split}
&p+\gamma\,\rightarrow\,\Delta\,\rightarrow\,p+\pi^{0}\\
&p+\gamma\,\rightarrow\,\Delta\,\rightarrow\,n+\pi^{+}
\end{split}
\end{equation}

The first process is dominant for protons with energy lower than the threshold energy $E=5\times10^{19}\,\text{eV}$, whereas the second process becomes the main attenuation mechanism for more energetic~protons.

The photopion production is the core mechanism of the GZK phenomenon~\cite{Greisen,Zatsepin} for protons. This~process gives an upper limit to the energy of CRs coming from distant sources, since through this effect a proton dissipates energy, but it is not annihilated.  If~the proton has enough energy, it can undergo the same interaction process again. This way, it becomes possible to evaluate the \emph{attenuation length} of a proton, defined as the average distance that the particle has to travel in order to reduce its energy by a factor of $1/e$. The~inverse of the attenuation length is given by:
\cite{Stecker1}:
\begin{equation}
\label{4}
\begin{split}
&\frac{1}{l_{p\gamma}}=\int_{\epsilon_{th}}^{+\infty}n(\epsilon)\,d\epsilon\int_{-1}^{+1}\frac{1}{2}\,s\,(1-v_{p}\,\mu)\,\sigma_{p\gamma}(s)\,K(s)\,d\mu=\\
&\phantom{\frac{1}{l_{p\gamma}}}=\int_{\epsilon_{th}}^{+\infty}n(\epsilon)\,d\epsilon\int_{-1}^{+1}\frac{1}{2}\,s\,(1-v_{p}\cos{\theta})\,\sigma_{p\gamma}(s)\,K(s)\,d\cos{\theta}
\end{split}
\end{equation}
where $\sigma_{p\gamma}(s)$ is the cross section for the proton--photon interaction as a function of the squared energy in the Mandelstam $s$ variable (the squared center of mass energy), $\mu$ is the impact parameter $\mu=\cos{\theta}$, $n(\epsilon)$ is the density of background photons per unit volume and photon energy $\epsilon$ and $\epsilon_{th}$ is the threshold energy for the interaction. $K(s)$ is the reaction inelasticity, defined as the energy fraction available for the production of secondary particles during the reaction. $K=(1-\eta)$ and $\eta={E_{\text{out}}}/{E_{\text{in}}}$ is the elasticity, defined as the energy fraction preserved by the primary particle after the interaction, with~$E_{\text{in}}$ the incoming particle energy and $E_{\text{out}}$ the residual energy. Using the following relations:
\begin{equation}
\label{5}
\begin{split}
&s=(m_{p}+\epsilon')^2-\left|\vec{p}_{\gamma}'\right|^2=m_{p}^2+2m_{p}\epsilon'\\
&\epsilon'=\gamma\epsilon(1-v_{p}\cos{\theta})
\end{split}
\end{equation}
where the Mandelstam $s$ is computed introducing the photon four momentum $(\epsilon',\,\vec{p}_{\gamma}')$ defined in the rest frame of the nucleus. In~the high energy limit approximation, it is possible to consider the proton velocity $v_{p}\simeq1$ and with $ds=-2E_{p}\epsilon\,d\cos{\theta}$ the Equation~(\ref{4}) becomes:
\begin{equation}
\label{6}
\frac{1}{l_{p\gamma}}=\frac{1}{2\,\gamma^{2}}\int_{\epsilon'_{th}/2\gamma}^{+\infty}d\epsilon \frac{n(\epsilon)}{\epsilon^2}\int_{\epsilon'_{th}}^{\epsilon'_{max}=2\gamma\epsilon}\frac{1}{2}\,\epsilon'\,\sigma_{p\gamma}(\epsilon')\,K(\epsilon')\,d\epsilon'
\end{equation}
where all the standard quantities are defined in the laboratory rest frame and the primed ones are defined in the proton rest frame. Using the fact that the background most relevant to UHECR propagation is the CMB, whose distribution $n(\epsilon)$ is a black-body spectrum with $T \approx 2.7~\text{K}$, the~previous relation (\ref{6}) can be simplified in the following form, obtaining the inverse of the attenuation length~\cite{Stecker}:
\begin{equation}
\label{7}
\frac{1}{l_{p\gamma}}=-\frac{k_{\text{B}}\,T}{2\,\pi^2\,\gamma^2}\int_{\epsilon'_{th}}^{+\infty}\epsilon'\,\sigma_{p\gamma}(\epsilon')\,K(\epsilon')\,\ln{\left(1-e^{-\epsilon'/2k_{\text{B}}T\gamma}\right)}\,d\epsilon'
\end{equation}

In the classical Lorentz invariant physics scenario, the inelasticity is given by the relation~\cite{Stecker1}:
\begin{equation}
\label{8}
K(s)=\frac{1}{2}\left(1-\frac{m_{p}^{2}-m_{\pi}^{2}}{s}\right)
\end{equation}


\section{LIV Introduced Phenomenology in UHECR~Propagation}\label{sec:propagation}
The LIV modifications introduced by the HMSR model are conceived in order to modify the kinematics, avoiding the introduction of exotic particles or reactions. The~kinematical perturbations modify the allowed phase space for the reactions and may therefore influence the processes involved in UHECR propagation, such as the GZK effect. The~introduction of LIV can indeed affect the photopion production, the~core reaction underlying the GZK phenomenon for~protons.

\subsection{$\Delta$ Resonance Creation~Constraint}
The photopion production passes mainly through a $\Delta$ particle creation (\ref{1}). The~process can occur passing through a real $\Delta$ particle (dominant process) or a virtual one. The~production of the real $\Delta$ particle must be preserved in order to foresee only small LIV-induced deviations from the classic GZK process. Indeed, if the real $\Delta$ particle production is kinematically forbidden, the photopion production effect and, as a consequence, even the GZK cutoff are strongly suppressed by the LIV introduction. The~free energy---that is, the squared root of the Mandelstam $s$ variable---of the proton interacting with the CMB photon must be larger than the rest energy of the $\Delta$ resonance, in~order to produce one. This~threshold energy consideration allows us to pose a first constraint on the LIV magnitude. Using~the modified internal product (\ref{a25}), the Mandelstam $s$ variable can be written as a function of the proton and the photon four momenta, respectively, $(E_{p},\,\vec{p}_{p})$ and $(\epsilon,\,\vec{p}_{\gamma})$:
\begin{equation}
\label{b1}
\begin{split}
&s=(E_{p}+\epsilon)^2-(\vec{p}_{p}[e^{-1}_{p}]+\vec{p}_{\gamma})^2\geq m_{\Delta}^{2}\;\Rightarrow\\
\Rightarrow\;&E_{p}^{2}-\vec{p}_{p}(1-f_{\mathbf{p}}(p_{p}))-2f_{\mathbf{p}}(p_{p})\vec{p}_{p}+2E_{p}\epsilon+\\
&-2\vec{p}_{p}\cdot\vec{p}_{\gamma}\left(1+\frac{1}{2}f_{\mathbf{p}}(p_{p})\right)\geq m_{\Delta}^{2}\\
\end{split}
\end{equation}
where the approximation $[e^{-1}_{p}](\mathbf{p})=1+\frac{1}{2}f_{p}(\mathbf{p})$ has been used for the vierbein (\ref{a14}), and $f_{p}(\mathbf{p})$ represents the proton LIV perturbation function. From~the previous relation (\ref{b1}), it is possible to derive the following relation:
\begin{equation}
\label{b2}
2f_{p}(p_{p})E_{p}^{2}-E_{p}(4\epsilon+f_{\mathbf{p}}(p_{p})\epsilon)+m_{\Delta}^{2}-m_{p}^{2}\leq0
\end{equation}

This second grade inequality must be satisfied in order to produce a $\Delta$ resonance; otherwise, the GZK effect is suppressed by the introduction of LIV. It is therefore simple to derive the following~constraint:
\begin{equation}
\label{b3}
f_{p}(p_{p})<\frac{\Delta M^2-4\epsilon-\sqrt{(\Delta M)^2-8E_{\gamma}\Delta M^2}}{4E_{\gamma}^2}
\end{equation}
where $\Delta M^2=(m_{\Delta}^{2}-m_{p}^{2})$. Admitting the average values $E_{\gamma}\simeq 7.0\times 10^{-4}\;\text{eV}$, $m_{\Delta}\simeq 1232\;\text{MeV}$, $m_{p}\simeq 938\;\text{MeV}$, we obtain the disequality:
\begin{equation}
\label{b4}
f_{\mathbf{p}}(p_{p})< 6\cdot10^{-23}
\end{equation}
to guarantee the existence of the GZK effect, comparable with the superior limit $4.5\times 10^{-23}$ obtained numerically in~\cite{Stecker} assuming that the observed cutoff is entirely due to the propagation. (Scenarios in which the sources themselves have an injection cutoff can be compatible with even stronger LIV in which GZK interactions are disabled altogether~\cite{Boncioli2,Boncioli,Lang}.)

\subsection{Reduced Phase Space and Modified~Inelasticity}
The introduction of LIV determines a change of the phase space allowed for the photopion reaction, determining a modification of the inelasticity $K$ (\ref{8}). The~modified inelasticity can be evaluated starting from the modified kinematics. Following the approach of~\cite{Stecker,Scully}, in order to simplify the computation, it is useful to define the center of momenta reference frame, where the following relation is satisfied:
\begin{equation}
\label{b5}
\vec{p}_{p}^{*}+\vec{p}_{\pi}^{*}=0
\end{equation}

In this last equation, the vectors are defined in $(TM,\,\eta_{ab})$. The~$\gamma_{CM}$ factor, involved in the change of reference frame from the center of momenta to a generic one, can be computed considering the free energy of the photo-pion production:
\begin{equation}
\label{b6}
\sqrt{s}=(E_{p}^{*}+E_{\pi}^{*})
\end{equation}
obtaining the equation:
\begin{equation}
\label{b7}
\gamma_{CM}(E_{p}^{*}+E_{\pi}^{*})=\gamma_{CM}\sqrt{s}=(E_{p}+E_{\pi})\;\Rightarrow\;\gamma_{CM}=\frac{E_{p}+E_{\pi}}{\sqrt{s}}=\frac{E_{tot}}{\sqrt{s}}
\end{equation}

The free energy necessary for the creation of a photo-pion in the CM frame of reference can be computed, using the CM definition (\ref{b5}) $\left|\vec{p}^{*}_{\smash{p}}\right|=\left|\vec{p}^{*}_{\pi}\right|$:
\begin{equation}
\label{b8}
\begin{split}
&(\sqrt{s}-E^{*}_{p})^2-([e_{\pi}^{-1}]\vec{p}^{*}_{p})^2=m_{\pi}^{2}\Rightarrow \\
\Rightarrow&(s-2\sqrt{s}E^{*}_{p})+E_{p}^{*2}-\left|\vec{p}^{*}_{\smash{p}}\right|^{2}(1+f_{p})+\left|\vec{p}^{*}_{\smash{p}}\right|^{2}f_{p}-\left|\vec{p}^{*}_{\smash{p}}\right|^{2}f_{\pi}=m_{\pi}^{2}
\end{split}
\end{equation}
where $f_{p}$ and $f_{\pi}$ represent respectively the proton and pion LIV correction~functions.

From the previous equation follows the equality:
\begin{equation}
\label{b9}
E_{p}^{*}=\frac{s+m_{p}^2+f_{p}\left|\vec{p}^{*}_{\smash{p}}\right|^{2}-m_{\pi}^{2}-f_{\pi}\left|\vec{p}^{*}_{\smash{p}}\right|^{2}}{2\sqrt{s}}=\frac{s+m_{p}^2-m_{\pi}^2+f_{p\pi}\left|\vec{p}^{*}\right|^{2}}{2\sqrt{s}}=F(s)
\end{equation}
where $f_{p\pi}=f_{p}-f_{\pi}$ is the LIV~parameter.

From this relation, it emerges immediately that the residual proton energy after the photopion reaction is increased by the proton correction and reduced by the pion one. This means that the proton correction must be larger than the pion one in order to produce a GZK sphere enlargement. In~this work, we make the assumption that every particle has a maximum attainable velocity lower than that of light $c$. Moreover, we admit that the heavier particles have a bigger LIV induced modification, since~the supposed quantum space perturbations have a gravitational origin. This allows us to consider the correction factor of the pion negligible and to pose $f_{p\pi}\simeq f_{p}$.

Now, it is possible to use the approximations:
\begin{equation}
\label{b10}
\begin{split}
&p^{*}_{p}\simeq E^{*}_{p}=(1-k_{\pi}(\theta))\sqrt{s}\\
&p^{*}_{\pi}\simeq E^{*}_{\pi}=k_{\pi}(\theta)\sqrt{s}
\end{split}
\end{equation}

In these equations, $\sqrt{s}$ represents the initial free total energy and $E'_{p}$ and $E'_{\pi}$ are the final energies of the proton and the pion,~respectively.

At this point using the Lorentz changing frame equation, as~a result, it is possible to write:
\begin{equation}
\label{b11}
\begin{split}
&E'_{p}=\gamma_{CM}(E_{p}^{*}+\beta \cos{\theta} p^{*}_{p})\\
&E^{*}_{p}=(1-k_{\pi}(\theta))\sqrt{s}
\end{split}
\end{equation}
using the pion~inelasticity.

Approximating the three-momentum magnitude with the energy and the velocity factor $\beta$ with~1, in~the hypothesis of ultra-relativistic particles and substituting $\gamma_{CM}$ with the value computed in Equation~(\ref{b7}), it is possible to obtain the following equation:
\begin{equation}
\label{b12}
(1-k_{\pi}(\theta) )=\frac{1}{\sqrt{s}}\left(F(s)+\cos{\theta}\sqrt{F(s)^2-m_{p}^2+2f_{p}\left|\vec{p}\right|^2}\right)
\end{equation}
from which it is possible to compute the inelasticity in function of the collision angle $\theta$. The~equation is solved introducing the energy of the photon $\epsilon'$ defined in the reference frame where the proton momentum is null $\vec{p}_{p}=0$, so the $s$ free energy can be written as:
\begin{equation}
\label{b12a}
s=(E_{p}+\epsilon')^2-\left|\vec{p}_{\gamma}\right|^2=2m_{p}\epsilon'+m_{p}^2
\end{equation}

We can now average the inelasticity on the interval $\theta\in[0,\;\pi]$:
\begin{equation}
\label{b13}
k_{\pi}=\frac{1}{\pi}\int_{0}^{\pi}k_{\pi}(\theta)\;d\theta
\end{equation}

In Figures~\ref{fig:inel1}--\ref{fig:inel3}, we plot the inelasticity computed in the case of different LIV parameter \mbox{$f_{p\pi}\simeq f_{p}$} values as a function of the interacting proton $E_{p}$ and photon $\epsilon$ energies in the laboratory rest frame. In~Figure~\ref{fig:inel1}, we report the inelasticity obtained in a Lorentz invariant scenario and given by Equation~(\ref{8})---LIV parameter posed equal to $f_{p\pi}\simeq f_{p}=0$. In~Figure~\ref{fig:inel2}, we plot the inelasticity computed in the case of a LIV parameter $f_{p\pi}\simeq f_{p}=9\times10^{-23}$. Finally, in Figure~\ref{fig:inel3}, the inelasticity is plotted for $f_{p\pi}\simeq f_{p}=3\times10^{-24}$. The~LIV caused effects determining a dramatic drop in the inelasticity value (shown in the $E_{p}-\epsilon'$ space where the proton energy $E_{p}$ is defined in the laboratory reference frame and the photon one $\epsilon'$ is defined in the proton rest frame) for the different values of the LIV parameter. This implies a drastic reduction of the allowed phase space for the creation of a photopion during the interaction of a proton with a CMB photon (\ref{b9}).

\begin{figure}[H]
\centering
\includegraphics[scale=0.32]{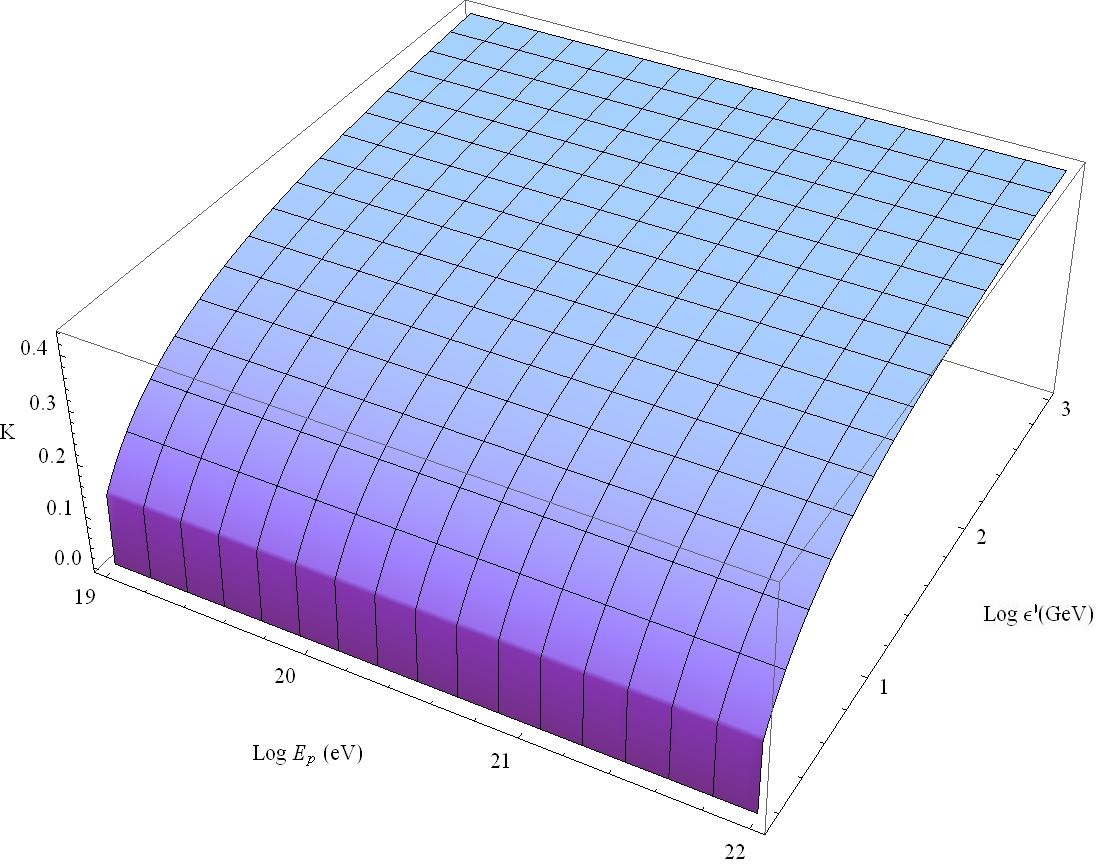}
\caption{\textit{Cont}.}
\label{fig:inel1}
\end{figure}
\begin{figure}[H]
\centering

\includegraphics[scale=0.30]{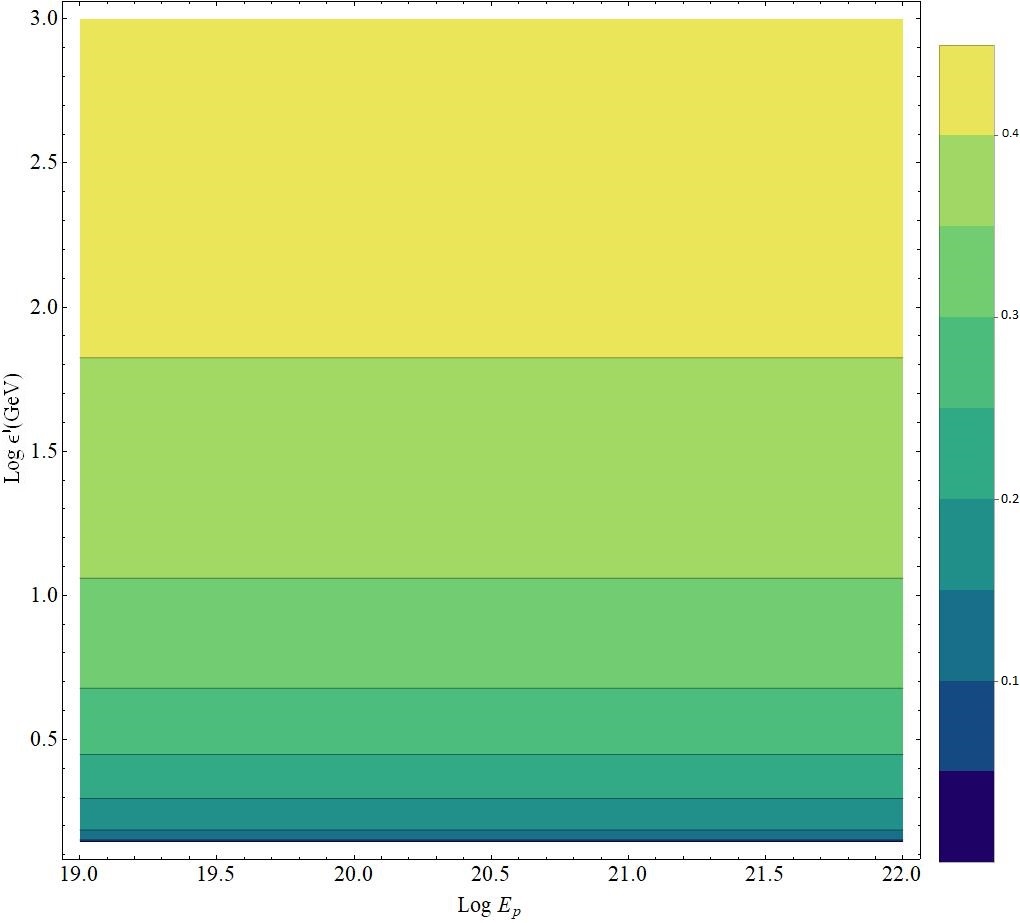}
\caption{Inelasticity obtained in the case of LIV parameter $f_{p\pi}\simeq f_{p}=0$ as a function of the proton energy $E_{p}$ and of the photon energy $\epsilon'$ defined in the proton rest~frame.}
\label{fig:inel1}
\end{figure}
\unskip

\begin{figure}[H]
\centering
\includegraphics[scale=0.35]{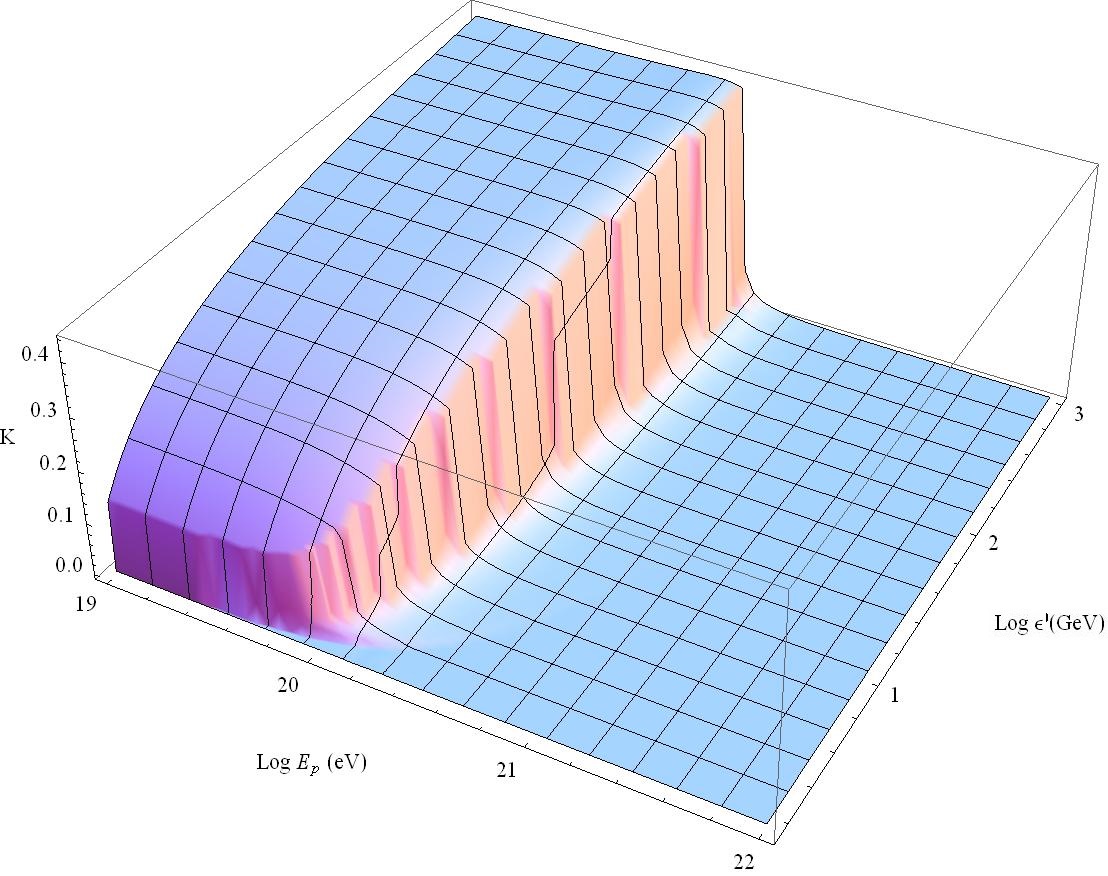}
\caption{\textit{Cont}.}
\label{fig:inel2}
\end{figure}
\begin{figure}[H]
\centering
\includegraphics[scale=0.30]{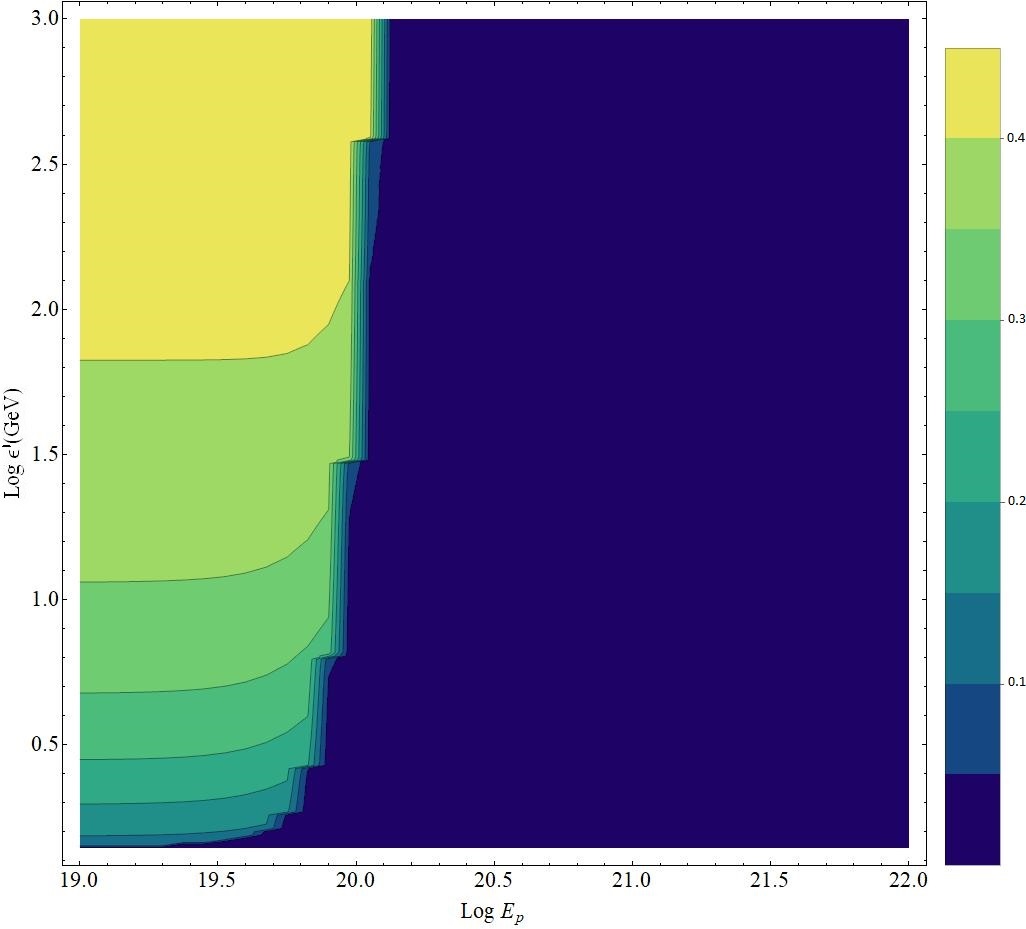}
\caption{Inelasticity obtained in the case of LIV parameter $f_{p\pi}\simeq f_{p}=9\times10^{-23}$ as a function of the proton energy $E_{p}$ and of the photon energy $\epsilon'$ defined in the proton rest~frame.}
\label{fig:inel2}
\end{figure}
\unskip

\begin{figure}[H]
\centering
\includegraphics[scale=0.35]{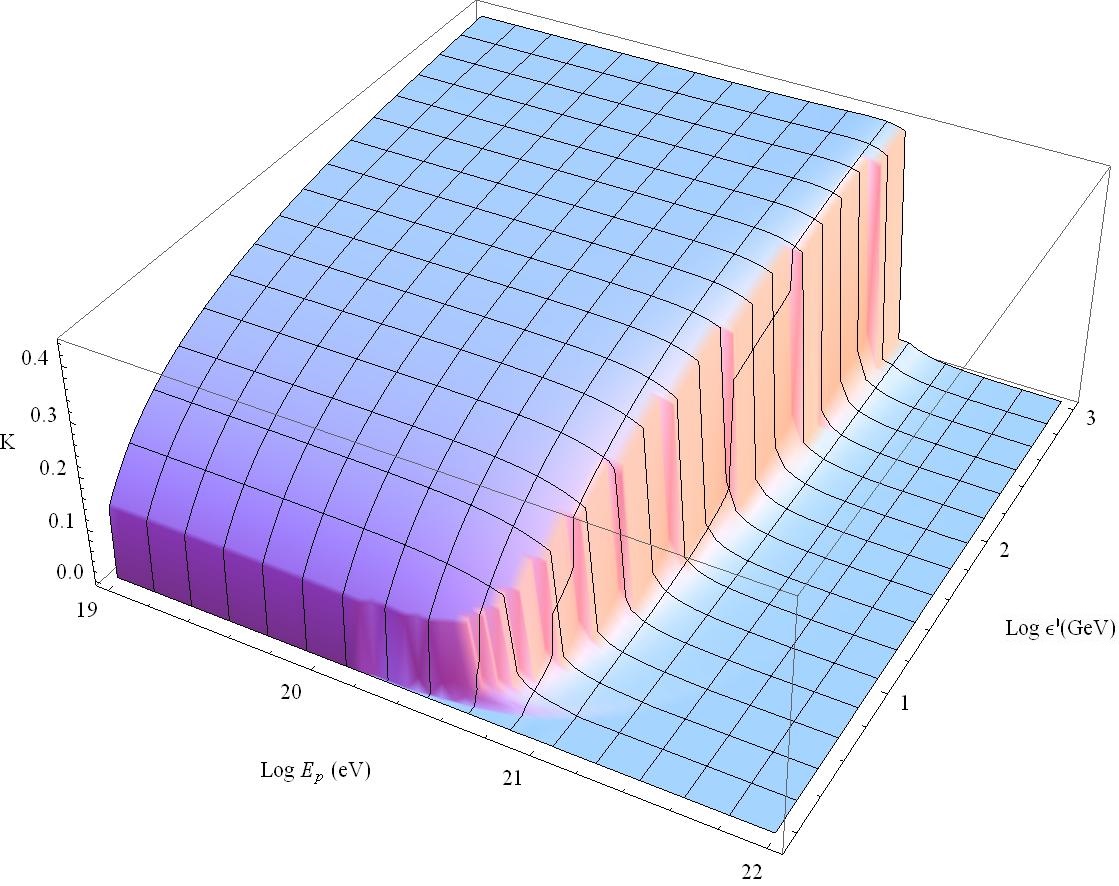}
\caption{\textit{Cont}.}
\label{fig:inel3}
\end{figure}
\begin{figure}[H]
\centering
\includegraphics[scale=0.30]{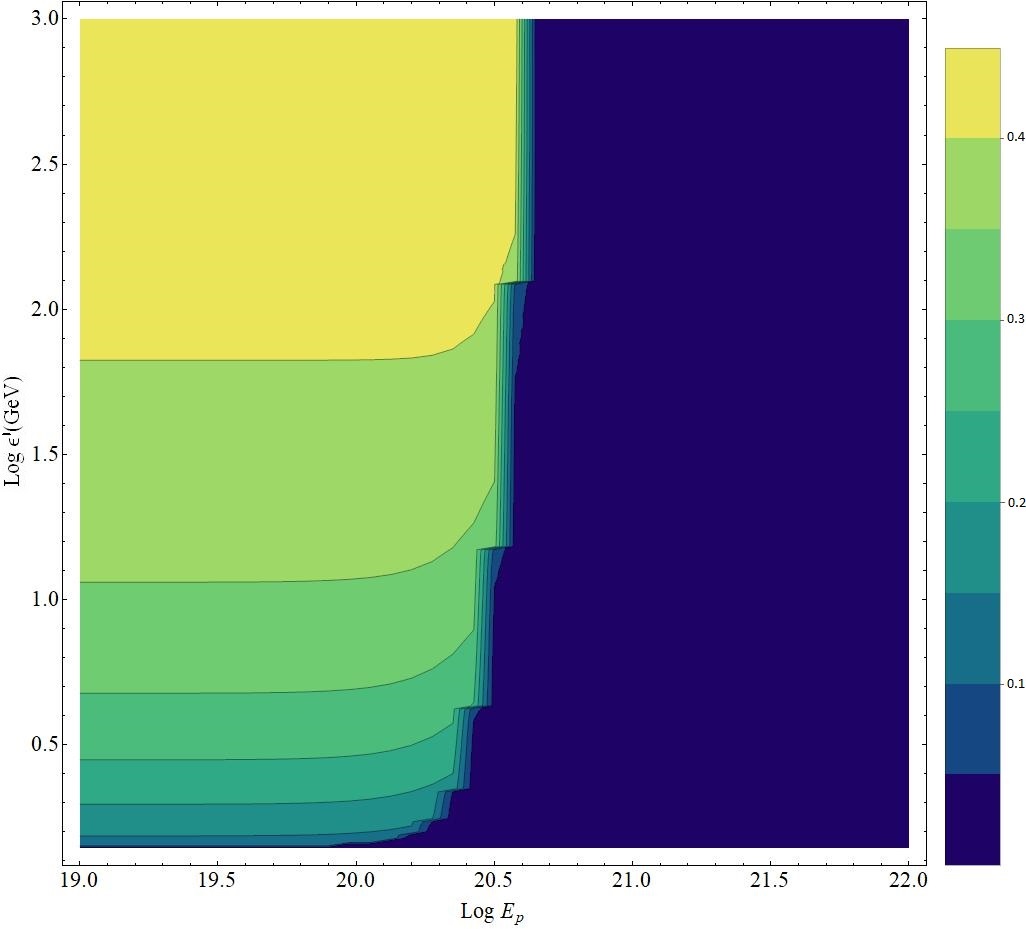}
\caption{Inelasticity obtained in the case of LIV parameter $f_{p\pi}\simeq f_{p}=3\times10^{-24}$ as a function of the proton energy $E_{p}$ and of the photon energy $\epsilon'$ defined in the proton rest~frame.}
\label{fig:inel3}
\end{figure}

\section{Simulations on LIV Modified~Propagation}\label{sec:simulations}

Making use of the modified shape of the inelasticity $K$ function derived from the theory, we can now compute the modifications experienced by an UHECR proton for the GZK opacity horizon and the attenuation length, assuming that their production rate is constant in space and time.  To~define the GZK opacity sphere, we consider the region in which $50\%$ of the protons that reach Earth with a residual energy higher (or equal to) the threshold of the photopion production process, which is $5\times10^{19}\; \text{eV}$, has~originated.

In order to evaluate the modifications introduced by the LIV theory, we simulated the UHECR protons' free propagation using a specifically modified version of the simulation program \textit{SimProp} \cite{Aloisio}. In~the modified version of the software, we substituted the classical physics predicted inelasticity with the amended one given by an analytic function of the energy of the proton, the~CMB photon interacting energy defined in the proton rest frame and of the LIV parameter. This modification determines that the GZK produced photopion energy is quickly reduced in the case the energies of the interacting particles belong to a determined range in the phase space, since the value of the inelasticity in Figures \ref{fig:inel1} to \ref{fig:inel3} drops quickly to negligible~values.

\begin{figure}[H]
\centering
\includegraphics[scale=0.75]{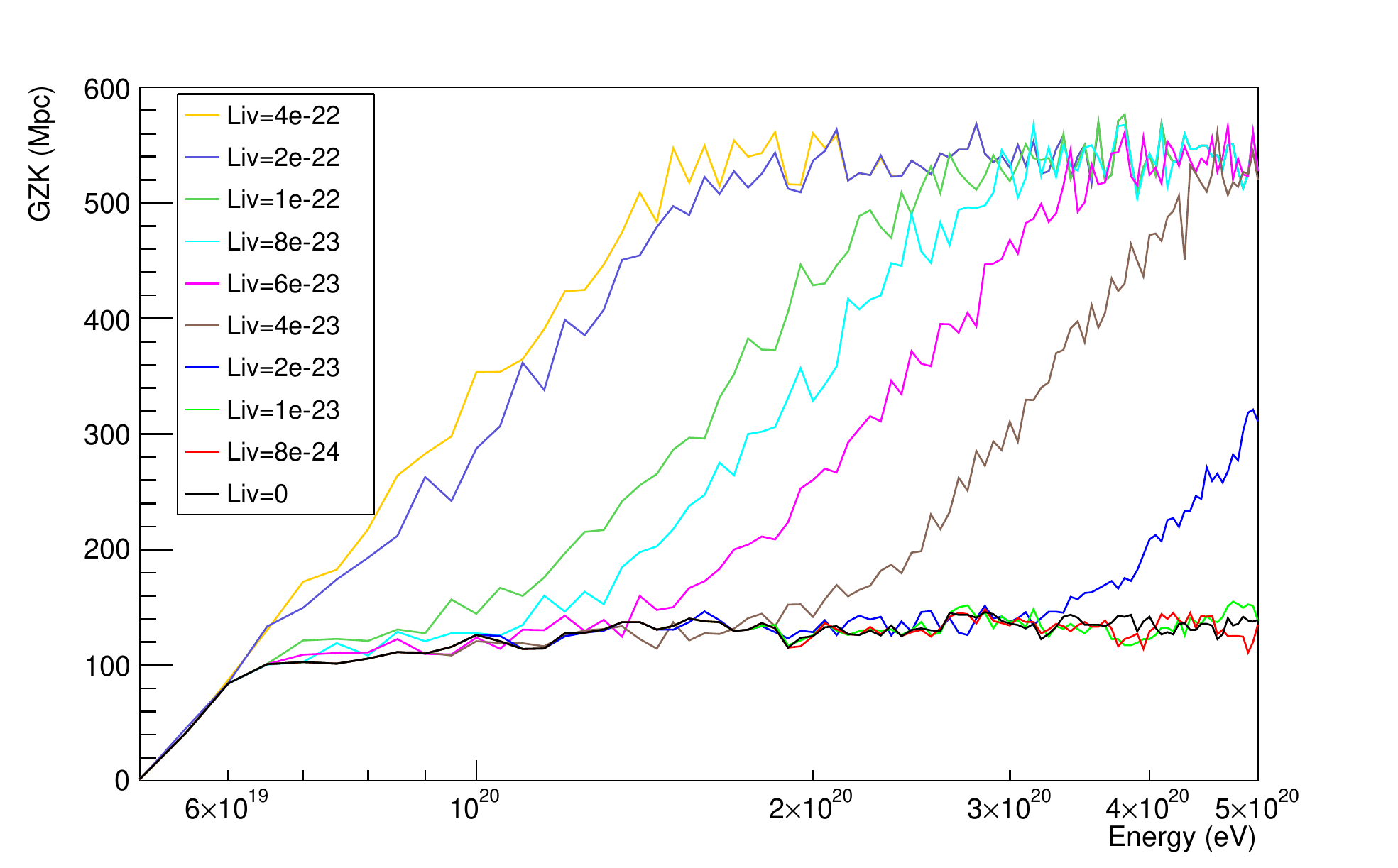}
\caption GZK sphere radius as a function of production energy $E$ simulated for $1000$ protons generated randomly inside of a sphere with a radius of $\sim$$1270\, \text{Mpc}$, centered on Earth. The~simulations for ten different values of the LIV parameter are shown in the energy range $0\,\text{eV}$--$2\times10^{20}\,\text{eV}$.
\label{fig:GZK}
\end{figure}

The LIV-caused GZK sphere modifications are evaluated simulating the propagation of 1000~protons for various LIV parameters magnitudes with various fixed energies. The~results of the simulations are then used to plot the GZK opacity sphere radius as a function of the energy and the LIV parameters (see Figure~\ref{fig:GZK}). Every plot line is obtained fixing the LIV parameter difference magnitude from a minimum value of $8\times10^{-24}$ to a maximum of $4\times10^{-22}$ and is compared with the result obtained for null LIV. For~fixed LIV parameters, the 1000 particles are initially simulated with an acceleration energy $E=5\times10^{19}\,\text{eV}$. The~simulations are then repeated increasing the production energy by steps of $5\times10^{18}\,\text{eV}$ reaching a maximum value of $E=2\times10^{20}\,\text{eV}$, in~order to obtain every plot line. We made the conservative assumption that, in a CR group with varying production energy, the dominant component is the lowest one. Indeed, the energy loss rate increases steeply with increasing CR energy, so it is possible to simplify the analysis considering every detected proton as generated with the inferior energy value for every CR group. The~particles are simulated assuming a uniform distribution inside a sphere of radius $\sim1270\, \text{Mpc}$ centered on Earth that is inside a portion of Universe included between redshift parameters $z=0$ and $z=0.25$.

\begin{figure}[H]
\centering
\includegraphics[scale=0.5]{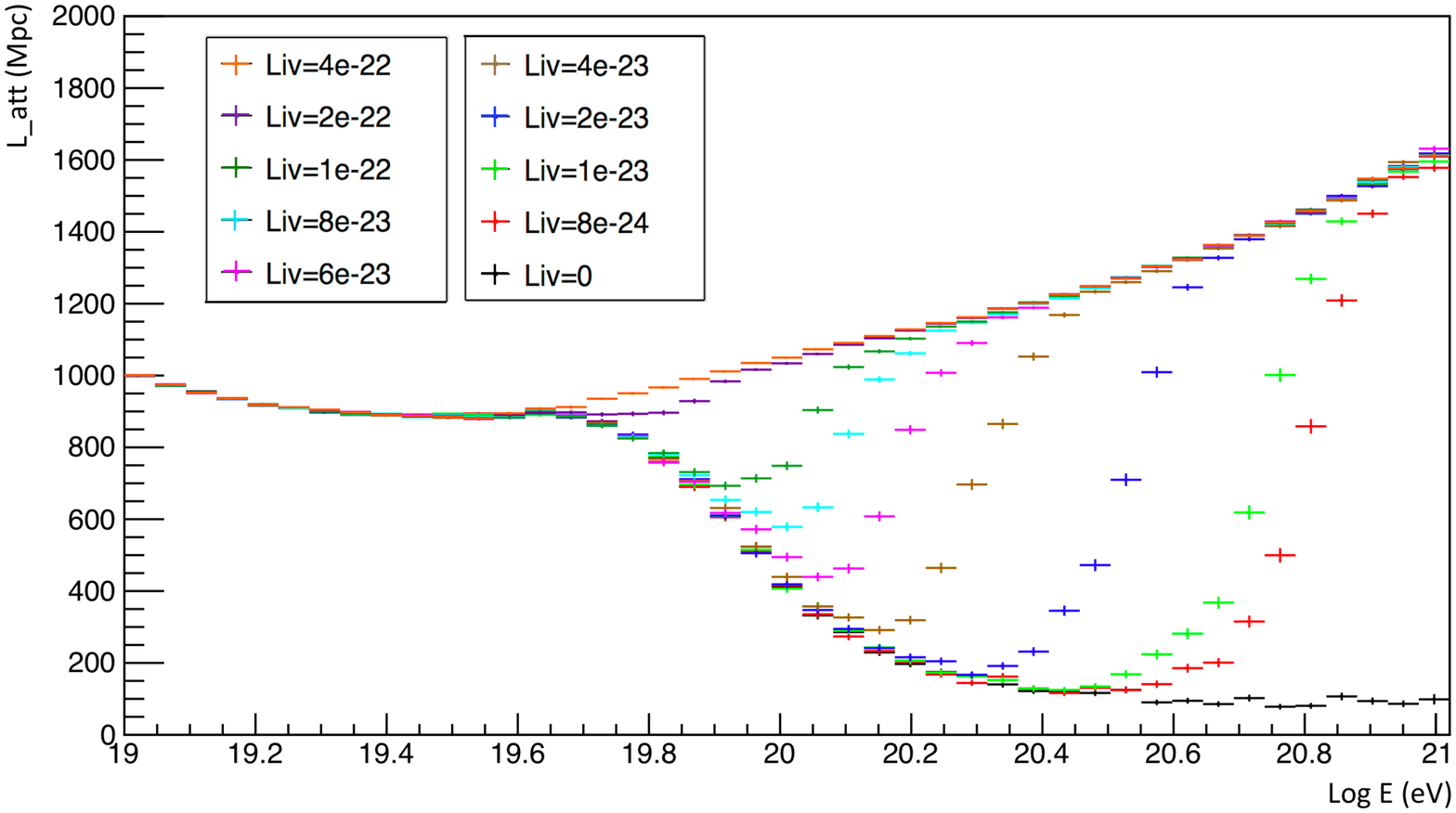}
\caption{ Attenuation length as a function of energy $E$, plotted for ten different values of the LIV~parameter.}
\label{fig:Latt}
\end{figure}

From Figure~\ref{fig:GZK}, one can clearly notice how the opacity sphere for UHECR protons is modified by the LIV introduction; indeed, the GZK proton horizon is LIV parameter dependent. Figure~\ref{fig:GZK} has visible inflections points where the steepness of the GZK radius plot increases as a function of the energy. These inflection points depend on the LIV parameter magnitude; indeed, they are present at lower energies for increasing LIV values. This behaviour is correlated with analogous inflections that are present in the plot of the attenuation length (Figure \ref{fig:Latt}) as a function of energy and LIV as explained~below.
Following the same simulation procedure, we can also plot the attenuation length for the reaction. The~results of the simulations for different LIV parameters are shown in Figure~\ref{fig:Latt}. In~the LIV-less scenario, the attenuation length decreases with increasing energy values. In~the presence of LIV perturbations, it is clear that the simulated energy loss length undergoes a recovery that becomes more and more important for increasing values of the LIV parameter. Indeed, the attenuation length at first decreases with increasing energy; then, there is an inflection point and, after this, the increases for increasing energy.  This effect is caused by the average energy lost in every photopion production process that is reduced by LIV. Indeed, up to LIV parameters of $10^{-22}$, the variation from the Lorentz invariant (LI) case is noticeable starting from production energy values of $\sim$$10^{20}\,\text{eV}$, while, for LIV parameters at least equal to $2\times10^{-22}$, the modification is appreciable starting already from $\sim$$6\times10^{19}\,\text{eV}$. The~inflection points depends therefore on the LIV magnitude as in the case of GZK-cutoff plot. Note that we are neglecting the effects of LIV on electron--positron pair production interactions, which for large values of the LIV parameter would presumably further increase the energy loss~lengths.
As a result, one can state that the GZK sphere enlargement and the attenuation length increase imply that the introduction of LIV determines a reduction of the Universe opacity for UHECR~propagation.

\section{Conclusions}\label{sec:conclusions}
Lorentz covariance underlies nowadays our knowledge of physics; nevertheless, small departures from this fundamental symmetry are supposed as possible residual evidence of the phenomenology induced by quantum gravity. Our study is conducted in the framework of HMSR theory a model that foresees a LIV minimal extension of the SM of particle physics preserving the covariance of the formulation in a context able to predict a rich phenomenology. HMSR theory foresees a minimal extension of the SM of particle physics that preserves the usual gauge invariance $SU(3)\times SU(2)\times U(1)$ and does not introduce exotic particles or reactions. Moreover, the theory preserves the covariance in an amended formulation that does not require the introduction of a preferred reference frame. This last aspect is of great experimental advantage since all the obtained observations concerning UHECR can be collected without discriminations related to the orientation of the observatory respect to a privileged reference frame. This model can therefore be used to conduct studies of correlation between the anisotropy of the arrival direction of UHECR with the distribution of candidate sources thanks to the preservation of the spacetime~isotropy.
Our study is conducted by simulating the propagation of UHECR protons in the presence of LIV using an ad~hoc modified version of the SimProp software. The~obtained result foresees an enlargement of the GZK opacity sphere and of the correlated attenuation lengths. This predictions are in accordance with other works~\cite{Stecker,Scully,TorriUHECR,Bietenholz:2008ni,Mattingly:2005re}, but,~with the difference of being obtained in a covariant preserving framework. The~accordance of these results with previous works indicates a great accuracy of the \emph{SimProp} modifications introduced to take into account the LIV generated effects on UHECR protons and guarantees a solid basis for the research framework introduced~here.
The theoretical predictions about the $\Delta$ resonance creation suggest a LIV parameter value of $\sim$$6\times10^{-23}$ (\ref{b4}), similar to values found in previous works~\cite{Stecker,Kostelecky:2008ts} by fitting the observed UHECR spectrum assuming a pure proton injection with no cutoff at sources. To~obtain concrete results from the comparison with experimental data about UHECR, it may be necessary to improve our study. The~new physics introduced by LIV leaves room for an analysis refinement studying a more realistic scenario. Future developments could implement the study of heavier UHECRs (for instance, the intermediate component), in~order to follow the indications given by the Pierre Auger Collaboration that privileges a mixed UHECR composition~\cite{Aab:2014aea}. Indeed, LIV effects are expected to be larger for intermediate nuclei due to the nature of the kinematical modifications, which could affect their propagation in a more relevant way than that of the lighter or heavier UHECR components. We made an excessive conservative hypothesis considering only protons that is the lighter UHECR component, but,~to support this first approximation, it is important to underline that all the results obtained for this scenario are still valid for the more realistic one, for~which they are expected to be even more important. The~results reported in this paper can be used to compare predictions with experimental observations. In~particular, one may try and compute how observables such as spectrum, composition, and~secondary photon flux are expected to change at different LIV intensities, as~was done in~\cite{Boncioli2,Diaz,Boncioli,Klinkhamer,Lang2,Lang}. Moreover, since~the HMSR has the peculiarity of preserving spacetime isotropy, it is particularly fit to make predictions on how UHECR anisotropy evolves with LIV in different sources scenarios, as~we plan to do this in a forthcoming~work.

\vspace{30pt}



\textbf{Author contributions}\\
Writing-original draft: M.D.C.T., L.C., A.d.M., A.M. and L.M.; review and editing: M.D.C.T.. All authors
contributed equally to this work. All authors have read and agreed to the published version of the manuscript. 

\textbf{Funding}\\
This work was partially supported by a post doctoral fellowship financed by the Fondazione Fratelli Giuseppe Vitaliano, Tullio e Mario Confalonieri, Milano Italy.


\textbf{Conflicts of interest}\\
The authors declare no conflict of interest.

\textbf{Abbreviations}\\
The following abbreviations are used in this manuscript:
\begin{itemize}
  \item \textbf{UHECR} Ultra High Energy Cosmic Ray
  \item \textbf{CR} Cosmic Rays
  \item \textbf{LIV} Lorentz Invariance Violation
  \item \textbf{HMSR} Homogeneously Modified Special Relativity
  \item \textbf{SM} Standard Model
  \item \textbf{SME} Standard Model extension
  \item \textbf{VSR} Very Special Relativity
  \item \textbf{DR} Dispersion Relation
  \item \textbf{MDR} Modified Dispersion Relation
  \item \textbf{MLT} Modified Lorentz Transformation
\end{itemize}


\appendix
\section{Derivative of the 0 Degree Homogeneous~Function} \label{app}
One fundamental feature of HMSR is the possibility to establish a direct relation between the momentum and the coordinate space, at~least at the leading order. This peculiar characteristic follows from the possibility to neglect the derivatives of the co-metric by momentum. The~negligibility of these derivatives implies that the geometric structure is asymptotically flat (that is approximately flat but not trivial) and the total covariant derivative can be approximated with the standard partial one. All~the previous considerations follow from the homogeneity of the perturbation function $f$ in the MDR~(\ref{a1}).
The homogeneity of the perturbation function $f$ is fundamental to neglect the derivatives by the momentum, which is proportional to terms with the form:
{\small
\begin{equation}
\label{c17}
\frac{\partial}{\partial p_{\mu}}f\left(\frac{|\vec{p}|}{E}\right)
\end{equation}}

Finally, taking the Taylor expansion series of the $f$ function and the high energy limit, it follows the relation:
{\small
\begin{equation}
\label{c18}
\begin{split}
&\partial_{p^{j}}f(p)=\partial_{p^{j}}\sum_{k}\alpha_{k}\frac{|\vec{p}|^{k}}{E^{k}}\simeq\partial_{p^{j}}\sum_{k}\alpha_{k}\frac{|\vec{p}|^{k}}{(\sqrt{|\vec{p}|^2+m^2})^k}=\\
=&\sum_{k}\left(\alpha_{k}k\frac{|\vec{p}|^{k-2}p_{j}}{(\sqrt{|\vec{p}|^2+m^2})^k}-\alpha_{k}k\frac{|\vec{p}|^{k}p_{j}}{(\sqrt{|\vec{p}|^2+m^2})^{k+2}}\right)\rightarrow0
\end{split}
\end{equation}}

\noindent where the approximate equivalence $E\simeq\sqrt{|\vec{p}|^2+m^2}$ has been~used.

\end{document}